\documentclass[11pt,letterpaper]{article}
\pdfoutput=1
\usepackage{soul}
\usepackage{jheppub}
\usepackage{amsfonts}
\usepackage{amsmath}
\allowdisplaybreaks[4]     
\usepackage{amssymb}   
\usepackage{euscript}       
\usepackage{color}
\usepackage[normalem]{ulem}
\usepackage{mathrsfs}  
\usepackage{booktabs}
\usepackage{siunitx}
\usepackage{amsmath,latexsym,xcolor,mathtools}
\usepackage{graphicx,verbatim, subcaption, caption,inputenc}

\newcommand{\bea}{\begin{eqnarray}}
\newcommand{\eea}{\end{eqnarray}}
\newcommand{\ba}{\begin{eqnarray}}
\usepackage{braket}
\newcommand{\ea}{\end{eqnarray}}

\newcommand{\beq}{\begin{equation}}
\newcommand{\eeq}{\end{equation} }
\newcommand{\beqa}{\begin{eqnarray}}

\newcommand{\eeqa}{\end{eqnarray}}
\newcommand{\beqar}{\begin{eqnarray*}}
\newcommand{\eeqar}{\end{eqnarray*}}

\newcommand{\be}{\begin{equation}}
\newcommand{\ee}{\end{equation}}

\begin{document}

\title{CFT correlators from shape deformations in Cubic Curvature Gravity}
\author[a,b]{Giorgos Anastasiou,}
\author[a,b]{Ignacio J. Araya,} 
\author[a,b,c]{Andrés Argandoña}
\author[d]{and Rodrigo Olea}

\emailAdd{ganastasiou@unap.cl}
\emailAdd{ignaraya@unap.cl}
\emailAdd{aargandona@estudiantesunap.cl}
\emailAdd{rodrigo.olea@unab.cl}
\affiliation[a]{Instituto de Ciencias Exactas y Naturales (ICEN), Universidad Arturo Prat, Avenida Arturo Prat Chac\'on 2120, 1110939, Iquique, Chile}
\affiliation[b]{Facultad de Ciencias, Universidad Arturo Prat, Avenida Arturo Prat Chac\'on 2120, 1110939, Iquique, Chile}
\affiliation[c]{Departamento de Física de Altas Energías, Instituto de Ciencias Nucleares,
Universidad Nacional Autónoma de México,
Apartado Postal 70-543, CDMX 04510, México}
\affiliation[d]{Departamento de Ciencias F\'{\i}sicas, Universidad Andres Bello, Sazi\'{e} 2212, Piso 7, Santiago, Chile}

\abstract{
We find a covariant expression for the universal part of the holographic entanglement entropy which is valid for CFTs dual to generic higher curvature gravities in up to five bulk dimensions. We use this functional to compute universal coefficients of stress-tensor correlators in three-dimensional CFTs dual to Cubic Curvature Gravity. Using gauge/gravity duality, we work out an expression for the entanglement entropy of deformed entangling regions and read the coefficients from the power expansion of the entropy in the deformation parameter. In particular, we obtain the $t_4$ coefficient of the 3-point function and exhibit a difference between the results obtained using the entanglement entropy functional for minimal and non-minimal splittings. We compare the obtained expressions for $t_{4}$ derived considering both splittings with results obtained through other holographic methods which are splitting-independent. We find agreement with the result obtained from the non-minimal splitting, whereas the result derived from the minimal splitting is inconsistent and it is therefore ruled out. 
}
 
\maketitle

\section{Introduction}

It is known that the fixed points of the renormalization group (RG) flows of quantum field theories (QFTs) are described by scale invariant theories which, in most cases, turn out to have conformal symmetry \cite{Polyakov:1970xd,Polchinski:1987dy,Nakayama:2010zz}. Therefore, critical phenomena can be  modelled by conformal field theories (CFTs). They characterize phase transitions in several physical systems such as the second order phase transition of water \cite{stanley1971phase} and the ferromagnetic-paramagnetic transition evidenced in the Ising model \cite{onsager1944two}. In addition, the study of CFTs might lead to a better understanding of the theory space of QFTs, since there are conformal fixed points at the ends of their RG flows. These CFTs are characterized by their central charges, which count the number of degrees of freedom and the behaviour of their correlation functions.

In the context of the gauge/gravity duality \cite{Maldacena:1998,Gubser:1998bc,Witten:1998qj}, the $d$-dimensional CFTs can be described in terms of their quantum partition function, which in the saddle-point approximation is given by the on-shell gravity action of the corresponding $D=(d+1)$-dimensional dual bulk theory. The interest on the study of higher-curvature gravity (HCG) theories
lies in their ability to describe more general CFTs than the ones that have Einstein-AdS as their gravity dual \cite{Henningson:1998gx,Henningson:1998ey,Nojiri:1999mh}. Among the properties of these CFTs, of particular importance are the coefficients of the contact-term expansion of their stress-tensor correlators. These coefficients are part of the definition of the theory and have different interpretation. For instance, $C_{\text{T}}$ is the coefficient of the two-point function, which directly controls the unitarity of the theory, subject to a positivity requirement \cite{Osborn:1993cr}. Also, $C_{\text{T}}$, $t_\text{2}$ and $t_\text{4}$ are coefficients which characterize the three-point function and control the energy flux that reaches an observer located in a specific direction at null infinity \cite{Osborn:1993cr,Erdmenger:1996yc,Hofman:2008ar}.  

The AdS/CFT correspondence provides us the tools to express the aforementioned quantities in geometrical terms. In particular, the fact that entanglement entropy (EE) can be used as an order parameter to characterize quantum phase transitions and critical points in different phases of matter \cite{Kitaev:2005dm, Klebanov:2007ws}, makes it the ideal probe of the central charges of the corresponding CFT. The introduction of a purely geometrical object for the computation of EE \cite{Ryu:2006bv}, in the same spirit as the Bekenstein-Hawking entropy for the black hole, is the key for the geometrical interpretation of the CFT universal coefficients.

More specifically, Ryu and Takayanagi (RT) propose a novel  way  to compute the entanglement entropy of a given spatial region $A$ for a CFT that is holographically dual to Einstein gravity. In this framework, the EE is proportional to the area of a bulk codimension two surface $\Sigma$, which is the minimal one among the surfaces homologous to the spatial boundary region $A$,  anchored at $\partial \Sigma$ \footnote{$\partial \Sigma$ is conformally equivalent to the entangling surface $\partial A$.}, i.e.,

\begin{equation}
    S(A)=\frac{Area(\Sigma)}{4G}. \label{RTformula}
\end{equation}
This formula resembles the one given by Bekenstein and Hawking for the computation of the black hole entropy, where the horizon is the fixed point set of a global $U(1)$ symmetry. Later on, Lewkowycz and Maldacena (LM) in Ref.\cite{Lewkowycz:2013nqa} generalized the Gibbons and Hawking method \cite{Gibbons:1976ue}  for the calculation of  gravitational entropy to cases without $U(1)$ symmetry. In the context of the AdS/CFT duality, the generalized gravitational entropy, after taking the tensionless limit, corresponds to the von Neumann entropy of a well-behaved density matrix in the CFT side. In the case of Einstein gravity, the prescription given by LM constitutes the proof of the  conjecture stated by the equation \eqref{RTformula}. 

The generalization of gravitational entropy beyond $U(1)$ symmetry, allowed to extend the standard replica trick, used in QFT, to the bulk. On the field theory side, the replica method is particularly useful for the calculation of Renyi entropy, expressing it in terms of the partition function evaluated on an $n$-fold cover $\mathcal{M}_n$, which in turn is obtained by cyclically gluing $n$ copies of the original space $\mathcal{M}$ along the entangling region ($A$) of interest, with an explicit $\mathbb{Z}_n$ symmetry \cite{Calabrese:2004eu,Holzhey:1994we}. 

In the holographic context, the dictionary dictates that there is a bulk space $\mathcal{B}_n$, dual to the boundary $\mathcal{M}_n$. Therefore, the bulk manifold features an explicit replica symmetry. In particular, the action of the $\mathbb{Z}_n$ symmetry induces an orbifold structure in the bulk of the form $\hat{\mathcal{B}}_n=\mathcal{B}_n/\mathbb{Z}_n$ that is regular everywhere except at the fixed points, where a conical singularity appears with deficit angle $2\pi(1-1/n)$. Using the GKP-W relation \cite{Gubser:1998bc,Witten:1998qj}, in the saddle point approximation, the Renyi entropy is given by the formula

\begin{equation}
  S_n=\frac{n}{n-1}(I[\hat{\mathcal{B}}_n]-I[\mathcal{B}_1]) .
\end{equation}
The EE is obtained as the $n=1$ limit of the above equation, which becomes $S(A)=\lim_{n\rightarrow 1} S_n=\partial_n{I}[\hat{\mathcal{B}}_n]|_{n=1}$. Therefore, the problem of finding the EE of a given region in the CFT has been translated into a problem in classical gravity.

The generalization of the RT formula for higher curvature theories of gravity was worked out in Refs.\cite{Dong:2013qoa,Camps:2013zua}, based on the LM prescription. Following the notation of Ref.\cite{Dong:2013qoa}, this is achieved, by the introduction of a set of adapted coordinates on $\Sigma$ that describes the near conical singularity geometry as

\begin{equation}
    ds^{2}= e^{2 \Omega} \left(d\rho^{2}+\rho^{2} d\tau^{2}\right)+ \left(\sigma_{ij}+2 K_{Aij}x^{A}\right) dy^{i} dy^{j} + \ldots \,,
    \label{metric_adapted}
\end{equation}
where $\left(\rho,\tau\right)$ represent the two-dimensional orthogonal space to $\Sigma$ and $\sigma_{ij}$ is the induced metric given by the  embedding function $x^{\mu}=x^\mu(y^i)$. In general, one may replace $\left(\rho,\tau\right)$ with the holomorphic coordinates $(z,\bar{z})$ through the mapping $z=\rho e^{i \tau}$, such that the warp factor $\Omega$ with a thickness parameter $b$ reads $\Omega=-\frac{1}{2} \left(1-\frac{1}{n}\right) \log\left(z \bar{z}+b^2\right)$. In this frame, the two  normal vectors to $\Sigma$ are $n^1=\sqrt{\frac{z}{\bar{z}}}\partial_z+\sqrt{\frac{\bar{z}}{z}}\partial_{\bar{z}}$ and $n^2=i\left(\sqrt{\frac{z}{\bar{z}}}\partial_z-\sqrt{\frac{\bar{z}}{z}}\partial_{\bar{z}}\right)$. Finally, the tensor $K^{A}_{ij}$ of the above metric, where $A=1,2$, is the extrinsic curvature of the surface $\Sigma$ along the normal direction $(n^A)_{\nu}$ and it is defined as $K^A_{i j}=e^{\mu}_{i}e^{\nu}_{j}\nabla_\mu (n^A)_{\nu}$, where $e^{\mu}_{i}=\partial_ix^\mu$ are tangent vectors. The extrinsic curvature can be expressed in a covariant form as $K^{\mu}_{\nu \rho}=K^{A}_{i j} (n_A)^{\mu}e_{\nu}^{i}e_{\rho}^{j}$.

For the general Euclidean higher curvature  Lagrangian\footnote{Only contractions of the Riemann tensor are considered. For the case in which covariant derivatives are taken into account, see ref.\cite{Miao:2014nxa}.}  $\mathcal{L}_{\text{HCG}}(R_{\rho \sigma}^{\mu\nu})$, the EE adopts the form

\begin{equation}
S_{\text{HCG}}=\frac{1}{8G}\int\limits_{\Sigma} d^{d-1}y\sqrt{|\sigma|}\left[\frac{\partial\mathcal{L}_{\text{HCG}}}{\partial R_{z\bar{z}z\bar{z}}}+\sum_{\alpha}\left(\frac{\partial^2\mathcal{L}_{\text{HCG}}}{\partial R_{zizj}\partial R_{\bar{z}k\bar{z}l}}\right)_{\alpha}\frac{8K_{zij}K_{\bar{z}kl}}{q_{\alpha}+1}\right] \,. \label{Dong}
\end{equation}
Notice that this expression consists of two parts; the first term corresponds to the Wald's entropy formula \cite{Wald:1993nt,Jacobson:1993vj}; and the second is the \textit{anomalous term} coming from the potentially logarithmic divergent contributions in the action at the conical singularity when taking the $n=1$ limit. 
 
The main issue in the above expression is that different ways of regularizing the action near the conical singularity give rise to different EE functionals. This ambiguity arises in the anomalous term of the Dong's formula, being known in the literature as the \textit{splitting problem} \cite{Camps:2014voa,Miao:2014nxa,Miao:2015iba}. There, it becomes clear that different regularizations of the conical apex lead to different universal characteristics for the CFT in question\footnote{For more details on the splitting problem, see Appendix \ref{Appendix A}.}. Point in fact, Ref.\cite{Miao:2015iba} discusses that using the splitting given by Ref.\cite{Dong:2013qoa} (minimal prescription) yields an inconsistent result for the universal terms of the EE for 6d CFTs and therefore a different regularization is desirable. 
 
Among the various regularization prescriptions, the requirement for the regularized metric, written in adapted coordinates at the conical singularity, to satisfy the EOM to the leading order seems the most prominent. Implementing this constraint for a general HCG theory is a difficult task. However, assuming that the theory is linearizable, to impose Einstein EOMs is enough. This so called non-minimal prescription, modifies the terms present in the expansion of the EE functional in terms of the curvatures in the adapted coordinates of Eq.\eqref{Dong}. This leads to a different anomalous terms than the one that was originally proposed. This procedure reproduces the correct universal terms for EE, at least perturbatively \cite{Camps:2014voa,Miao:2015iba,Camps:2016gfs}.

Even though any HCG theory is subject to the splitting problem, when deriving the corresponding holographic  entanglement entropy (HEE) functional there are certain examples free of these ambiguities. In particular, both prescriptions agree on the resulting HEE functionals of quadratic curvature gravity (QCG) \cite{Fursaev:2013fta} and Lovelock gravity \cite{Hung:2011xb,Camps:2013zua,deBoer:2011wk}. In the case of QCG this is understood, since the second order derivative with respect to the Riemann curvature gives a trivial contribution, independent of it. As for the case of Lovelock gravity, it is the universality of topological invariants and their dimensional continuations in the presence of conical singularities what makes them independent of the splitting problem. That imposes strict constraints on the corresponding EE functional. Indeed, it can be shown that for both the minimal and non-minimal splittings, Eq.\eqref{Dong} reduces to the Jacobson-Myers entropy \cite{Jacobson:1993xs}.

Leaving aside these two examples, a discrepancy between the two splittings is the general rule when higher curvature terms are considered. Therefore, if we want to obtain the central charges and correlation coefficients from the universal part of the EE, this may pose a serious issue. In this paper, our analysis  focuses on cubic curvature gravity  (CCG) theory, which is the lowest order Lagrangian in the curvature  where the splitting problem leads to different EE functionals \cite{Caceres:2020jrf}.

In section \ref{Section 2}, we discuss on the renormalization scheme used in order to determine the universal part of the EE for both the minimal and non-minimal splittings. Furthermore, we compute the $C$-function candidates of the CFT dual to CCG by evaluation on a  ball-shaped entangling region in arbitrary dimension. This determines the type-$A$ anomaly for even dimensional CFT and the $F$  quantity for the odd dimensional case. We also compute the type-$B$ anomaly evaluating the entropy on a cylindrical entangling region in $d=4$. 

In section \ref{section 3}, the splitting problem becomes manifest in the computation of the renormalized EE of a deformed entangling surface. There, the universal EE is expanded up to quartic order in the deformation parameter. Then, the coefficients of the two- and three-point correlation functions of the stress-energy tensor, $C_{T}$  and  $t_4$, are read-out from said expansion. We compare the $t_4$ charge for the two splittings with its corresponding value given in the literature, derived in a splitting-independent way by different holographic techniques. Using the latter as the \emph{golden rule}, we conclude that the only consistent prescription is the non-minimal one.

\section{Universal EE in Cubic Curvature Gravity}\label{Section 2}

Our analysis focuses on cubic curvature theories of gravity and their respective HEE functionals. As it was pointed out in the previous section, the minimal and non-minimal prescriptions give a different entropy functional when higher curvature terms are considered. The explicit expressions for the two splittings \eqref{HEE in CCG minimal Appendix} and \eqref{HEE in CCG nonminimal Appendix}  were given in Ref.\cite{Caceres:2020jrf}, where it is evident that the difference between them is encoded in the term containing fourth order contractions of the extrinsic curvature \eqref{SK4 min}-\eqref{SK4 non-min}. In what follows, we will isolate the universal part of the EE for each of the two splittings by adding a counterterm $S_{\text{KT}}$. This term is inherited from the bulk renormalization procedure via the Kounterterm $B_{d}$ of Eqs.\eqref{KTsodd} and \eqref{KTseven}, taking advantage of its self-replication property \cite{Anastasiou:2019ldc}.

The Kounterterm scheme was originally introduced for Einstein-AdS gravity in Refs.\cite{Olea:2005gb,Olea:2006vd}. Several generalizations of this formalism, including extensions to quadratic gravity \cite{Giribet:2018hck,Giribet:2020aks} and Lovelock theory \cite{Kofinas:2007ns,Kofinas:2008ub}  have been explored. In Ref.\cite{Araya:2021atx} it was proven that this scheme also works for generic HCG theories with AdS asymptotics in lower or equal than five bulk dimensions. This method defines a well posed variational problem for holographic boundary conditions in Einstein-AdS gravity and it is  compatible with  the standard holographic renormalization scheme for asymptotically conformally flat (ACF) spacetimes \cite{Anastasiou:2020zwc}.

Based on these considerations, we will initially isolate the universal part of the HEE for CFTs dual to generic HCG theories and then we specialize to the CCG case.

\subsection{Renormalized entanglement entropy in generic higher curvature gravity}

The identification of the universal terms of the HEE functional for HCG theories requires the asymptotic analysis of the Dong functional in Eq.\eqref{Dong}. This is achieved by inserting the Fefferman-Graham (FG) expansion of the bulk metric in the aforementioned formula. Interestingly enough, the embedding function $x^{\mu}=x^\mu(y^i)$ acquires a similar asymptotic expansion, that allows us to track down the divergences of any functional located in $\Sigma$.

In particular, the extrinsic curvature along the normal directions to the bulk codimension-2 surface $\Sigma$, that appear in the anomalous term of Eq.\eqref{Dong}, is of order $\mathcal{O}(z)$ in the Poincare coordinate \cite{Anastasiou:2021swo}, independently of the particular shape of the extremal surface. This is a direct consequence of the universality of the $\mathcal{O}(z^2)$ term of the embedding of $\Sigma$ \cite{Schwimmer:2008yh} which in turn results from the universality of the second-order coefficient of the boundary metric $g_{(2)ij}$ \cite{Imbimbo:1999bj}. Thus, the anomalous term of the Dong functional contributes at order $\mathcal{O} \left(z^2\right)$, since terms quadratic in the extrinsic curvature are involved. Furthermore, in Ref.\cite{Araya:2021atx} it was shown that for a generic HCG Lagrangian, in $d\leq4$, the following relation holds

\begin{equation}
    \frac{\partial  \mathcal{L}}{\partial R^{\mu\nu}_{\rho \sigma}}=\left.\frac{\partial  \mathcal{L}}{\partial R^{\mu\nu}_{\rho \sigma}} \right|_{AdS}+\mathcal{O}(z^2) \,.
\end{equation}
Therefore, the HEE Lagrangian of a generic HCG theory evaluated on a manifold with AdS asymptotics is given by the Wald term evaluated in pure AdS, up to $O(z^2)$ contributions.\footnote{Note that this fact relies on the $g_{(1)ij}$ coefficient of the FG expansion being equal to zero, which is true for the generic HCG theory, but it is not true for a specific measure-zero subset of theories which includes New Massive Gravity in $D=3$ at the special point, Conformal Gravity in $D=4$ and Einstein-Gauss-Bonnet Gravity at the Chern-Simons point in $D=5$.} Thus, without loss of generality, we can write

\begin{equation}
S_{\text{HCG}}=\frac{1}{8G}\int\limits_{\Sigma} d^{d-1}y\sqrt{|\sigma|}\left[\left.\frac{\partial\mathcal{L}_{\text{HCG}}}{\partial R_{z\bar{z}z\bar{z}}}  \right|_{AdS}+\mathcal{O}(z^2)\right] \,.
\end{equation}
Noting that, in up to four boundary dimensions, the $\mathcal{O}(z^2)$ term does not contribute to the power-law divergences -although it can contribute to the universal part-, we obtain that

\begin{equation}
S_{\text{HCG}}=\frac{1}{8G}\int\limits_{\Sigma} d^{d-1}y\sqrt{|\sigma|}\left[\left.\frac{\partial\mathcal{L}_{\text{HCG}}}{\partial R_{z\bar{z}z\bar{z}}}  \right|_{AdS} \right]+(\text{Univ}) \,,
\end{equation}
where $(\text{Univ})$ may be an unspecified contribution to the universal term. Finally, the Wald part of the HEE functional evaluated on pure AdS is always proportional to contractions of an antisymmetric product of Kroenecker deltas \cite{Bueno:2020uxs}, making the term proportional to the area of $\Sigma$ up to the theory-dependent overall factor $C(\ell_{\mathrm{eff}},\vec{\mu})$, given in Eq.\eqref{HCGCoupling}; namely

\begin{equation}
S_{\text{HCG}}=\frac{C(\ell_{\mathrm{eff}},\vec{\mu})}{4G}\int\limits_{\Sigma} d^{d-1}y\sqrt{|\sigma|}+(\text{Univ}) \,.
\end{equation}
It is evident from the last expression that in $d \leq4$, the structure of power-law divergences of the previous formula for generic HCG theories coincides with the one of the area of the codimension-2 surface $\Sigma$.

Rendering the area functional finite requires the addition of surface terms that cancel the divergent terms according to the chosen renormalization scheme. The extension of holographic renormalization 
in the context of HEE can be found in Refs.\cite{Taylor:2016aoi,Taylor:2020uwf}. An alternative and more geometric prescription to renormalize codimension-2 functionals considering the Kounterterm, has been used to obtain the HEE in the case of Einstein gravity \cite{Anastasiou:2017xjr,Anastasiou:2018rla,Anastasiou:2019ldc}, QCG \cite{Anastasiou:2021swo} and Lovelock gravity \cite{Anastasiou:2021jcv}, where the divergences were successfully removed and the universal part was recovered for the ball-shaped and cylindrical entangling regions. The main advantage for the use of Kounterterms is their self-replicating property, when evaluated on conically-singular manifolds, giving rise to the codimension-two extrinsic counterterm $B_{d-2}$. 

Interestingly enough, adding this surface term on top of the area functional with a fixed coupling constant $c_d^{\text{EH}}\bigg\lfloor \frac{d+1}{2}\bigg\rfloor$, where

\begin{equation}
c_d^{\text{EH}}=
    \begin{cases}
      \frac{(-1)^{\frac{d+1}{2}}\ell_{\mathrm{eff}}^{d-1}}{(\frac{d+1}{2})(d-1)!},& \mbox{if $d$ is odd},\\
      \frac{(-1)^{\frac{d}{2}}\ell_{\mathrm{eff}}^{d-2}}{2^{d-3}d((\frac{d}{2}-1)!)^2}, & \mbox{if $d$ is even},\\
    \end{cases}   \label{coupling Einstein Hilbert}    
\end{equation}
is sufficient to cancel the power-law divergences of the latter in $d \leq 4$, as shown in Refs.\cite{Anastasiou:2018rla,Anastasiou:2021swo}. Based on these considerations, and adjusting the fixed coupling constant by taking into account the higher-curvature couplings of a generic HCG, the universal part of the corresponding HEE functional is given by

\begin{equation}
S_{\text{HCG}}^{\text{Univ}}= S_{\text{HCG}} + S_{\text{KT}}
\label{Univ_HEE_HCG}
\end{equation}
where

\begin{equation}
S_{\text{KT}}= \frac{c_d^{\text{HCG}}}{4G}\bigg\lfloor \frac{d+1}{2}\bigg\rfloor\int\limits_{\partial\Sigma}d^{d-2}x\sqrt{|\tilde{\sigma}|}\,B_{d-2} \,,
\label{HEE_HCG_functional}
\end{equation}
with $\tilde{\sigma}$ and $B_{d-2}$ being the determinant of the induced metric and the extrinsic counterterms defined on $\partial\Sigma$, respectively.

This expression allows us to extract universal information encoded in the EE of a CFT that is dual to a generic HCG theory. In what follows, we use this result for the specific case of CCG, which is the simplest theory where the splitting problem manifests.

\subsection{The Cubic Curvature case}

In principle, the determination of the HEE for any HCG is subject to the splitting problem. CCG contains the lowest order terms in the curvature where this ambiguity becomes manifest. The most generic cubic curvature bulk Lagrangian is given by

\begin{align}
    \mathcal{L_{\text{CCG}}}&=R-2\Lambda_0+\mu_1R_{\mu\:\:\nu}^{\:\:\rho\:\:\sigma}R_{\rho\:\:\sigma}^{\:\:\lambda\:\:\tau}R_{\lambda\:\:\tau}^{\:\:\mu\:\:\nu}+\mu_2R^{\mu\nu}_{\:\:\:\:\rho \sigma}R^{\rho \sigma}_{\:\:\:\:\lambda \tau}R^{\lambda\tau}_{\:\:\:\:\mu\nu}
        +\mu_3R^{\mu\nu\rho}_{\:\:\:\:\:\:\:\sigma}R_{\mu\nu\rho\tau}R^{\sigma\tau}
        \nonumber\\
        &+\mu_4R_{\mu\nu\rho\sigma}R^{\mu\nu\rho\sigma}R+\mu_5R^{\mu\rho}R^{\nu\sigma}R_{\mu\nu\rho\sigma}+\mu_6R_{\mu}^{\:\:\nu}R_{\nu}^{\:\:\rho}R_{\rho}^{\:\:\mu}+\mu_7R_{\mu\nu}R^{\mu\nu}R+\mu_8R^3 \,, \label{CubicLagrangian}
\end{align}
where the Einstein-Hilbert (EH) part has also been included. Here, $\Lambda_0=-\frac{d(d-1)}{2\ell_{0}^2}$ is the negative cosmological constant of AdS and $\ell_{0}$ is the corresponding AdS radius. Rendering the action finite amounts to the addition of the surface terms \eqref{KTsodd} and \eqref{KTseven} with an overall theory-dependent coupling that is fixed by requiring the renormalization of the vacuum solution. 
Thus, the renormalized CCG action is written as

\begin{equation}
    I_{\mathrm{CCG}}^{ren}=\frac{1}{16\pi G}\int\limits_{\mathcal{B}}d^{d+1}x\sqrt{|G|}\mathcal{L}_{\mathrm{CCG}}+\frac{c^{\mathrm{CCG}}_d}{16\pi G}\int\limits_{\partial \mathcal{B}}d^{d}x\sqrt{|h|} B_d \,,\label{RenormalizedCubicCurvatureAction}
\end{equation}
where $c_d^{\text{CCG}} =c_d^{\text{EH}}C(\ell_{\mathrm{eff}},\vec{\mu})$ and the theory-dependent coefficient is given by

\begin{align}
C(\ell_{\mathrm{eff}},\vec{\mu})=&1+\frac{3}{\ell^4_{\mathrm{eff}}}\left[\mu_1(d-1)+4\mu_2+2\mu_3d+2\mu_4d(d+1)+\mu_5d^2+\mu_6d^2 \nonumber \right.\\
& \left. +\mu_7d^2(d+1)
 +\mu_8d^2(d+1)^2\right]\,.\label{CouplingCCG}
\end{align}
For this theory, the universal HEE computation simply requires adding the codimension-2 Kounterterm

\begin{equation}
    S_{\mathrm{KT}}=\frac{c_d^{\text{CCG}}}{4G}\bigg\lfloor \frac{d+1}{2}\bigg\rfloor\int\limits_{\partial\Sigma}d^{d-2}x\sqrt{|\tilde{\sigma}|}\,B_{d-2}\,,\label{HEE in CCG Kounterterm}
\end{equation}
which is a particular case of Eq.\eqref{HEE_HCG_functional}, resulting, for the minimal and non-minimal prescriptions, in

 \begin{align}
     S^{\text{Univ, min}}_{\text{CCG}}&=S^{\text{min}}_{\text{CCG}}+S_{\text{KT}},\label{EE universal min}\\
     S^{\text{Univ, non-min}}_{\text{CCG}}&=S^{\text{non-min}}_{\text{CCG}}+S_{\text{KT}}.\label{EE universal non-min}
 \end{align}
Indeed, this renormalization scheme determines the Weyl anomaly of even dimensional CFTs, given by the coefficient of the logarithmic divergence of the EE. In particular, for a region $A$ of size $L$, this quantity would be $S^{\text{Univ}} \left[A\right]=c_0\ln(\frac{L}{\delta})$, such that 

\begin{equation}
    c_0=(-1)^{d/2+1}2A\chi[\partial \Sigma]-\sum_{i}C_i\left(\partial_n\int\limits_{\mathcal{M}_n}I_i\right) \,.
\end{equation}
Here, $\chi[\partial \Sigma]$ is the Euler characteristic of the entangling surface and $I_i$ are the local conformal invariants in $d$ dimensions. The coefficients $A$ and $C_i$ are the type-$A$ and type-$B$ central charges, respectively. If we consider a ball-shaped entangling region, only the type-$A$ central charge will survive. On the other hand, if we consider a cylinder, only the type-$B$ charge will contribute to the universal EE.

In general, the central charges characterize and classify the particular CFT under consideration. The type-$A$ central charge is monotonous under a RG flow, thus being a good measure of the effective degrees of freedom of the theory \cite{Myers:2010xs}. This was initially demonstrated in CFT$_2$ by Zamolodchikov \cite{Zamolodchikov:1986gt}, and later generalized to four dimensions in Ref.\cite{Komargodski:2011vj}. The extension of the $a$-theorem to odd-dimensional spacetimes is called the F-theorem and it states that the free energy of a CFT on a Euclidean sphere is also monotonous under the RG flow \cite{Klebanov:2011gs,Jafferis:2011zi}. This was proven in three dimensions by Casini, Huerta and Myers (CHM) in Ref.\cite{Casini:2011kv}, considering the entropic inequalities that the EE of a disc satisfies, whose finite term is proportional to the free energy of a CFT$_3$ on a sphere.

In what follows, we will extract the universal part of the EE for specific highly-symmetric shapes, in a CFT with a CCG bulk dual, based on the renormalization scheme presented in Eqs.\eqref{EE universal min} and \eqref{EE universal non-min}. In particular, we will compute the  type-$A$ central charge and the $F$ quantity for  even and odd $d$, respectively. In addition, we will calculate the EE for a cylindrical region in CFT$_4$ and finally determine the type-$B$ anomaly coefficient.



\subsection{Ball-shaped entangling region}

In order to isolate the universal information out of the CFT data, such as the central charges, it suffices to consider the vacuum state of the theory, whose gravity dual corresponds to pure AdS. For this reason, we write the AdS metric in Poincar\'e coordinates with Euclidean signature

\begin{equation}
    ds^2=\frac{\ell^2_{\text{eff}}}{z^2}\left(d\tau^2+dz^2+dr^2+r^2d\Omega_{d-2}^2\right) \,. \label{Poincare}
\end{equation}
The dual CFT resides at the conformal boundary located at $z=0$. As it can be seen from Eq.\eqref{Poincare}, the metric is divergent and a cutoff at $z=\delta$ must be introduced. This is identified with the ultraviolet cutoff in the CFT.

 The interior of the spherical entangling surface is parametrized as $r\leq R $ at constant time $\tau$. The latter extends into the bulk as a codimension two surface $\Sigma$ with embedding function $r=f(z)$. As shown in   Refs.\cite{ Bhattacharyya:2014yga, Caceres:2020jrf}, the embedding function that minimizes the  entropy functional in CCG (\ref{HEE in CCG minimal Appendix}),(\ref{HEE in CCG nonminimal Appendix}), is the hemisphere $r=\sqrt{R^2-z^2}$.
The induced metric reads

\begin{equation}
    ds_{\Sigma}^2=\sigma_{ij}\,dy^{i}dy^{j}=\frac{\ell^2_{\text{eff}}}{z^2}\left(\frac{R^2}{R^2-z^2} dz^2+(R^2-z^2)d\Omega_{d-2}^2\right) ,\label{induced metric}
\end{equation}
with normal vectors $n_{\mu}^{A}$

 \begin{equation}
     n^{1}_{\mu}=\left(\frac{\ell_{\text{eff}}}{z},0,0,...,0\right) \quad \text{and} \quad n^{2}_{\mu}=\left(0,\frac{\ell_{\text{eff}}}{\sqrt{r^2+z^2}},\frac{\ell_{\text{eff}}\,r}{z\sqrt{r^2+z^2}},...,0\right). \label{normal vector sphere}
 \end{equation}
The vanishing of the extrinsic curvature for this particular embedding, makes the minimal and non-minimal prescriptions to coincide, resulting in the HEE functional 
 
\begin{equation}
  S_{\text{CCG}} =\frac{\mathcal{A}[\Sigma]}{4 G}+\frac{1}{8 G}\int\limits_{\Sigma}d^{d-1}x\sqrt{|\sigma|}S_{\text{R}^2} \,,
\end{equation}
where $\mathcal{A}[\Sigma]$ is the area of the codimension-2 bulk surface and $S_{\text{R}^2}$ is defined in Eq.\eqref{SR2}. The last formula can equivalently be written as

\begin{equation}
    S_{\text{CCG}} = C(\ell_{\mathrm{eff}},\vec{\mu}) \frac{ \mathcal{A}[\Sigma]}{4 G}   \,.
\end{equation}
As a consequence, the universal part of the EE now reads

\begin{align}
     S_{\text{CCG}}^{\text{Univ}}&=\frac{C(\ell_{\mathrm{eff}},\vec{\mu})}{4 G}\left(\mathcal{A}[\Sigma]+c_d^{\text{EH}}\bigg\lfloor \frac{d+1}{2}\bigg\rfloor\int\limits_{\partial\Sigma}d^{d-2}x\sqrt{|\tilde{\sigma}|}B_{d-2}\right)\\
     &=\frac{C(\ell_{\mathrm{eff}},\vec{\mu})}{4G}\mathcal{A}^{\text{Univ}}[\Sigma]\ \,,
     \label{renormalized EE for a sphere}
\end{align}
with

\begin{equation}
\mathcal{A}^{\text{Univ}}[\Sigma]=
    \begin{cases}
      (-1)^{\frac{d-1}{2}}\frac{2^{d-1}\pi^{\frac{d-1}{2}}\ell_{\text{eff}}^{d-1}}{(d-1)!}\,, & \mbox{if $d$ is odd}\,,\\

      (-1)^{\frac{d}{2}-1}\frac{2\pi^{\frac{d-1}{2}}\ell_{\text{eff}}^{d-1}}{(\frac{d}{2}-1)!}\log{\left(\frac{2R}{\delta}\right)}\,, & \mbox{if $d$ is even}\,,\\
    \end{cases}   \label{Area universal}    
\end{equation}
being the renormalized area of the surface $\Sigma$. This corresponds to either the finite term of the area expansion for odd-dimensional boundaries or the logarithmic contribution for even-dimensional. In this particular geometry, Eq.\eqref{renormalized EE for a sphere} can be expressed in terms of the $\mathcal{C}$-function candidates $F$ and $A$ as

\begin{equation}
S_{\text{CCG}}^{\text{Univ}}[\Sigma]=
    \begin{cases}
      (-1)^{\frac{d-1}{2}}F & \mbox{if $d$ is odd \,,} \\
     
      (-1)^{\frac{d}{2}-1}4 A \log{\left(\frac{2R}{\delta}\right)}& \mbox{if $d$ is even,}\\
    \end{cases}   \label{Area universal}    
\end{equation}
with

\begin{equation}
    F=C(\ell_{\mathrm{eff}},\vec{\mu})\frac{2^{d}\pi^{\frac{d-1}{2}}\ell_{\text{eff}}^{d-1}}{8G(d-1)!}  \quad \text{and} \quad A=C(\ell_{\mathrm{eff}},\vec{\mu})\frac{\pi^{\frac{d-1}{2}}\ell_{\text{eff}}^{d-1}}{8G(\frac{d}{2}-1)!} \,, \label{anomaly}
\end{equation}
being the partition function of the CFT$_d$ evaluated on a sphere and the type-$A$ anomaly coefficient, respectively. These results are in complete accord with the corresponding $C$-function candidates given in Ref.\cite{Bueno:2020uxs}.

\subsection{Cylindrical entangling region}\label{Second test: Cylinder}

Similarly to what has been done in the previous section, we consider the 5D Wick-rotated AdS metric in Poincar\'e coordinates written as

\begin{equation}
    ds^2=\frac{\ell^2_{\text{eff}}}{z^2}\left(d\tau^2+dz^2+dx_3^2+dr^2+r^2d\phi^2\right) \label{Poincare cylinder} \,.
\end{equation}
In the case of a cylindrical entangling surface of radius $L$, the two leading contributions of the bulk embedding function $\Sigma$ are theory-independent \cite{Schwimmer:2008yh}, leading to the following expansion

\begin{equation}
     \Sigma:\left[\tau=const,  r=L\left(1-\frac{z^2}{4L^2}+\mathcal{O}(z^4)\right)\right] \,. \label{Entangling region bulk}
 \end{equation}
Thus, the induced metric adopts the form

\begin{equation}
    ds_{\Sigma}^2=\frac{\ell^2_{\text{eff}}}{z^2}\left[\left(1+\frac{z^2}{4L^2}+\mathcal{O}(z^4)\right)dz^2+dx_3^2+\left(L-\frac{z^2}{4L}+\mathcal{O}(z^4)\right)^2d\phi^2\right] \,,\label{induced metric}
\end{equation}
with normal vectors
 
 \begin{equation}
     n^{1}_{\mu}=\left(\frac{\ell_{\text{eff}}}{z},0,0,0,0\right)\quad \text {and} \quad n^{2}_{\mu}=\left(0,\frac{\ell_{\text{eff}}}{\sqrt{4L^2+z^2}},0,\frac{2\ell_{\text{eff}}L}{z\sqrt{4L^2+z^2}},0\right).\label{normal vectors cylinder}
 \end{equation}
Considering the geometric elements obtained in the equations above, defined by the embedding of the cylinder, the different terms in the HEE functionals of Eqs.\eqref{HEE in CCG minimal Appendix}-\eqref{HEE in CCG nonminimal Appendix} are explicitly evaluated and expanded in orders of the holographic coordinate. Doing this, one obtains: 

\begin{align}
     \frac{\mathcal{A}[\Sigma]}{4G} &=\frac{\pi H}{4l G}\ell^3_{\text{eff}}\left[\frac{L^2}{\delta^2}-\frac{1}{4}\ln\left(\frac{L}{\delta}\right)\right]\,,\label{area cylinder}\\
      S_{\text{R}^2}&=\frac{6}{\ell^4_{\text{eff}}}\left(3\mu_1+4\mu_2+8\mu_3+40\mu_4+16\mu_5+16\mu_6 +80\mu_7+400\mu_8\right)\,, \\
      S_{\text{K}^2\text{R}}&= \frac{z^2}{L^2\ell^4_{\text{eff}}}(40\mu_4+8\mu_3+12\mu_2-3\mu_1)+\mathcal{O}(z^4)\,,\\
      S_{\text{K}^4}^{\text{min}}&=S_{\text{K}^4}^{\text{non-min}}=\mathcal{O}(z^4). \label{SK4 for the cylinder}
\end{align}
Notice that for this embedding, the terms that contribute to the splitting problem ($S_{\text{K}^4}$), coincide up to normalizable order. In total, in the long cylinder approximation, the HEE functional for CCG reads
 
\begin{equation}
  S_{\text{CCG}}=\frac{\pi H}{4L G}\ell^3_{\text{eff}}\left[\frac{C (\ell_{\mathrm{eff}},\vec{\mu})L^2}{\delta^2}-\frac{1}{4} b_4\ln\left(\frac{L}{\delta}\right)\right]+\mathcal{O}(\delta)\,,\label{HEE cylinder}\\
\end{equation}
where $H$ is the length of the cylinder and

\begin{equation}
    b_4=1+\frac{1}{\ell^4_{\mathrm{eff}}}\left(21 \mu_\mathrm{1}-36 \mu\mathrm{2}-8 \mu_\mathrm{3}-40 \mu_\mathrm{4}+48 \mu_\mathrm{5}+48 \mu_\mathrm{6}+240 \mu_\mathrm{7}+1200 \mu_\mathrm{8}\right)\,,
    \label{b4}
\end{equation}
with $b_4$ being the coupling-dependent coefficient of the CFT$_4$ type-$B$ anomaly. Notice that in this embedding the type-$A$ anomaly vanishes identically, leaving $b_4$ as the sole CFT data that can be extracted. 
Furthermore, taking into account the value of the CCG coupling constant (\ref{coupling Einstein Hilbert}), the counterterm contribution is computed as

\begin{align}
 S_{\text{KT}}&=\frac{c_4^{\text{CCG}}}{4G}\bigg\lfloor\frac{4+1}{2}\bigg\rfloor\int\limits_{\partial\Sigma}d^2x\sqrt{|\tilde{\sigma}|}B_2 \nonumber\\
 &=-\frac{2c_4^{\text{CCG}}\pi H }{G}\frac{\ell_{\text{eff}}L}{\delta^{2}}+\mathcal{O}(\delta) \,.
\end{align}
As a consequence, the universal part adopts the form

\begin{equation}
     S_{\text{CCG}}^{\text{Univ}}=S_{\text{CCG}}+S_{\text{KT}}=-b_4\frac{\pi H\ell^3_{\text{eff}}}{16L G}\ln\left(\frac{L}{\delta}\right) \,.
\end{equation}
Finally, using that on general grounds the logarithmic contribution of the HEE obtains the form \cite{Hung:2011xb,Bhattacharyya:2013jma,Bhattacharyya:2014yga}

\begin{equation}
      S_{\text{HCG}}^{\text{Univ}}=-\frac{C_{1} H}{2L}\ln\left(\frac{L}{\delta}\right) \,,
\end{equation}
it is possible to isolate the type-$B$ anomaly coefficient $C_{1}$ of the corresponding holographic CFT$_4$ dual to CCG by simple comparison, obtaining  $C_{1}=b_4\frac{\pi\ell^3_{\text{eff}}}{8G}$.

\section{Correlation function coefficients from shape deformations}\label{section 3}

Correlation functions in QFT are obtained from functional variations of the partition function, when considering external sources. The AdS/CFT correspondence provides an explicit construction of the generating functional of the boundary field theory in terms of the gravity action \cite{Gubser:1998bc}. A direct analogy can be drawn for the case of the evaluation of the CFT partition function on the thermal sphere. In Refs.\cite{Bueno:2018yzo,Bueno:2020odt}, it was proven that the coefficients $C_{\text{T}}$ and $t_4$ can be derived precisely from the squashing of the thermal sphere for arbitrary CFTs, which comprise universality relations.  These parameters were computed independently in terms of the scaling dimension of the twist operator and the coefficient of the two-point correlation function of the displacement operators \cite{Chu:2016tps,Dong:2016wcf,Bianchi:2016xvf,Bueno:2018xqc}.

Following an alternative route, Mezei et al. \cite{Mezei:2014zla, Allais:2014ata} identified $C_\text{T}$ in the contributions to the universal EE of CFT$_3$ at quadratic order in $\epsilon$, when sinusoidal deformations parametrized by $\epsilon$ are induced in the spherical entangling surface. This suggests a relation between deformations of the entangling region and variations of the CFT partition function. Here, we extend this result, by providing the computation of both the $C_\text{T}$ and the $t_4$ charges, for an arbitrary CCG theory, entirely in terms of shape deformations of the entangling region. The interest on the analysis of this theory is mainly two-fold. On the first place, it is the simplest case where there might be a difference in the universal EE between the different splittings. Secondly, it provides the ground for the computation of a non-trivial $t_4$, unlike the case of Einstein-AdS.

\subsection{Review of the method}

The computation is performed by approximating the codimension-2 extremal surface with the usual RT surface, which provides a result that is valid perturbatively in the higher curvature couplings. We consider the EE evaluated for sinusoidally-deformed spherical entangling regions, expanded on the deformation parameter $\epsilon$, denoted as $\mathbb{S}_{\epsilon}$. The expansion is given by

\begin{equation}
    S^{\text{Univ}}_{\text{CCG}}(\mathbb{S}_{\epsilon})= S_{\text{CCG}}^{(0)}+\epsilon^2 S_{\text{CCG}}^{(2)}+\epsilon^4 S_{\text{CCG}}^{(4)}+\mathcal{O}(\epsilon^6) \,,
\end{equation}
where $S_{\text{CCG}}^{(0)}=S^{\text{Univ}}_{\text{CCG}}(\mathbb{S})$  is the renormalized EE of the unperturbed sphere and $S_{\text{CCG}}^{(2i)}$ correspond to the higher-order entanglement susceptibilities. For a generic CFT, $S^{(2)}$ is identified with the coefficient of the two-point correlation function of the stress-energy tensor, dubbed $C_{T} $ \cite{Faulkner:2015csl}. This is in accordance with earlier results in holography, given in Refs.\cite{Faulkner:2015csl,Fonda:2015nma,Carmi:2015dla,Witczak-Krempa:2018mqx,Nozaki:2013wia,Mezei:2014zla,Allais:2014ata}. Based on this result, we are expecting that the coefficients of higher $n$-point correlation functions will appear when terms beyond the quadratic order are involved. For example, the parameters $C_\text{T}$, $t_{2}$ and $t_4$ of the three-point function.

As a warm-up exercise we will initially work on the EH case and then generalize the method to the theory of interest (CCG). For this reason, we consider the Poincaré-AdS$_4$ spacetime

\begin{equation}
    ds^2=\frac{\ell_{0}^2}{z^2}\left(d\tau^2+dz^2+dr^2+r^2d\phi^2\right). \label{Poincare 2}
\end{equation}
Since we are interested in Einstein-AdS gravity, the  corresponding EE functional is obtained by using the RT formula \eqref{RTformula}.
We know  that the embedding function for the case of the RT surface is given in Ref.\cite{Allais:2014ata} and reads

\begin{equation}
    \Sigma_{\text{RT}}: \:\:r=\sqrt{1-z^2}\left(1+\epsilon \sum_{l}\left(\frac{1-z}{1+z}\right)^{l/2}\frac{1+lz}{1-z^2}(a_l\cos{(l\phi)}+b_l\sin{(l\phi}))+\mathcal{O}(\epsilon^2)\right).\label{embedding deformed sphere}
\end{equation}
By this procedure, and upon isolating the universal part, we get

\begin{equation}
    S_{\text{EH}}^{\text{Univ}}=-\frac{\pi \ell_{0}^2}{2G}\left(1+\epsilon^2\sum_{l}\frac{l(l^2-1)}{4}(a_l^2+b_l^2)+\mathcal{O}(\epsilon^4)\right).\label{EE deformed sphere for EH}
\end{equation}
Indeed, the subleading term of the formula \eqref{EE deformed sphere for EH}, can equivalently be written as

\begin{equation}
     S_{\text{EH}}^{\text{(2)}}=-\frac{\pi^4 C_{\text{T}}^{\text{EH}}}{24}\sum_{l}l(l^2-1)(a_l^2+b_l^2) \,,\label{SEE eps}
\end{equation}
where

\begin{equation}
C_{\text{T}}^{\text{EH}}=\frac{3\ell_{0}^2}{\pi^3G}
\label{CT_EH}
\end{equation}
is the coefficient of the two-point correlation function of the stress tensor.

\subsection{Deformed sphere I: $C_{\text{T}}$}

The use of the RT surface as the probe to compute the correlator coefficients, works not only for Einstein-AdS gravity, but for any HCG up to the perturbative order in the couplings \cite{Bueno:2020uxs}. Therefore, in what follows, we consider the CCG theory under study to extend the analysis of the previous subsection.

Since we are interested in the computation of $C^{\text{CCG}}_{\text{T}}$, it is sufficient to consider the deformation of the embedding of Eq.\eqref{embedding deformed sphere} up to order $\mathcal{O}\left(\epsilon\right)$, what results in an order $\mathcal{O}\left(\epsilon^2\right)$ term in the entropy functional. Thus, all the relevant quantities for the calculation of the EE in CCG \eqref{SR2}-\eqref{SK4 non-min}, adopt the following form:

\begin{align}
    S_{\text{R}^2}&=\frac{6 \left(2 \mu_\mathrm{1}+4 \mu_\mathrm{2}+6 \mu_\mathrm{3}+24 \mu_\mathrm{4}+9 \mu_\mathrm{5}+9 \mu_\mathrm{6}+36 \mu_\mathrm{7}+144 \mu_\mathrm{8}\right)}{\ell_{\text{eff}}^{4}}\,,\label{SR2 deformed CT}\\
    S_{\text{K}^2\text{R}}&= -\epsilon^2\frac{12z^4 (\mu_1 -4 \mu_2 -2 \mu_3 -8 \mu_4)}{\ell_{\text{eff}}^4(1-z^2)^2}\sum_{l}\left(\frac{1-z}{1+z }\right)^{l}(a_l^{2}+b_l^{2}) l^{2} (l^2 -1)^{2}+\mathcal{O}\! \left(\epsilon^{3}\right)\,,\\
    S_{\text{K}^4}^{\text{min}}&=S_{\text{K}^4}^{\text{non-min}}=\mathcal{O}\! \left(\epsilon^{4}\right)\,.\label{SK4 deformed CT}
\end{align}
The last line indicates that the splitting problem is irrelevant for the computation of the  $C^{\text{CCG}}_{\text{T}}$. As a consequence, the HEE functional can be written, up to quadratic order, as

\begin{align}
   S_{\text{CCG}}=&-\frac{C(\ell_{\mathrm{eff}},\vec{\mu})\pi \ell_{\text{eff}}^2}{2G}\left(1-\frac{1}{\delta}\right)-\epsilon^2\frac{\pi \ell_{\text{eff}}^2}{8G}\sum_l\bigg[(a_l^2+b_l^2)l(l^2-1)\bigg(1\label{HEE in CCG deformed sphere for CT}\\&+\frac{3}{\ell_{\text{eff}}^4}(4\mu_1-4\mu_2+2\mu_3+8\mu_4
   +9\mu_5+9\mu_6+36\mu_7+144\mu_8)\bigg)\nonumber\\
   &-\frac{C(\ell_{\mathrm{eff}},\vec{\mu})}{\delta}(a_l^2+b_l^2)l^2\bigg]\nonumber+\mathcal{O}(\epsilon^3).\nonumber 
\end{align}
The counterterm that renders the previous functional finite is given by the one-dimensional Chern form  $B_1=-\frac{2}{\ell_{\text{eff}}}+\mathcal{O}(\epsilon^3,\delta)$. Then, we replace it in Eq.\eqref{HEE in CCG Kounterterm}, with the coupling $c_{3}^{\text{CCG}}=\frac{C(\ell_{\mathrm{eff}},\vec{\mu})\ell_{\text{eff}}^2}{4}$, obtaining

\begin{equation}
  S_{\text{KT}}=  -\frac{\pi \ell_{\text{eff}}^2 C(\ell_{\mathrm{eff}},\vec{\mu})}{2G}\frac{1}{\delta}\left(1+\epsilon^2\sum_l\frac{l^2}{4}(a_l^2+b_l^2)\right)+\mathcal{O}(\epsilon^3,\delta) . 
\end{equation}
This last expression matches the divergent part of Eq.\eqref{HEE in CCG deformed sphere for CT}, such that adding them together ($S_{\text{CCG}}+S_{\text{KT}}$) we obtain the finite part of the EE as

\begin{align}
     S_{\text{CCG}}^{\text{Univ}}&=-\frac{\pi \ell_{\text{eff}}^2}{2G}\bigg[C(\ell_{\mathrm{eff}},\vec{\mu})+\epsilon^2\sum_l\frac{1}{4}(a_l^2+b_l^2)l^2(l^2-1)\bigg(1+\frac{3}{\ell_{\text{eff}}^4}(4\mu_1-4\mu_2+2\mu_3\nonumber\\&+8\mu_4+9\mu_5+9\mu_6+36\mu_7+144\mu_8)\bigg)\bigg] +\mathcal{O}(\epsilon^4).\label{EE in CCG universal for deformed sphere}
\end{align}
From the $\mathcal{O}\left(\epsilon^2\right)$ term of the above formula we can read off the $C_{\text{T}}^{\text{CCG}}$ for a cubic curvature gravity theory. Here, we consider that the polynomial on $l$ is the same for both EH and CCG as it depends only on geometric consideration and not on the particular theory. With that in mind, we compare the subleading terms in Eqs.\eqref{SEE eps} and \eqref{EE in CCG universal for deformed sphere}, noticing that

\begin{equation}
     S_{\text{CCG}}^{\text{(2)}}=\frac{\pi^4 C^{\text{CCG}}_{\text{T}}}{24}\sum_{l}l(l^2-1)(a_l^2+b_l^2),
\end{equation}
where

\begin{equation}
    C_{\text{T}}^{\text{CCG}}=\bigg(1+\frac{3}{\ell_{\text{eff}}^4}(4\mu_1-4\mu_2+2\mu_3+8\mu_4+9\mu_5+9\mu_6+36\mu_7+144\mu_8)\bigg)C^{\text{EH}}_{\text{T}}.\label{Ct CCG}
\end{equation}
Hence, the $C^{\text{CCG}}_{\text{T}}$ for CFTs dual to CCG theory is proportional to that of Einstein-AdS with a coupling-dependent coefficient. This result matches the one encountered in Ref.\cite{Bueno:2020uxs} but following a different computational method. 

\subsection{Deformed sphere II: $t_\text{4}$} \label{Deformed sphere II} \label{subsection 3.3}

Our analysis in the previous subsection, made manifest that the computation of $S_{\text{CCG}}^{\text{(2)}}$ does not involve any contributions from the splitting-dependent terms. Actually, Eq.\eqref{SK4 deformed CT} indicates that these terms should be expected in the next subleading order, i.e. $S_{\text{CCG}}^{\text{(4)}}$. In this case, the order $\mathcal{O} \left(\epsilon\right)$  terms of the embedding function \eqref{embedding deformed sphere} are not sufficient to determine $S_{\text{CCG}}^{\text{(4)}}$, and higher orders should be considered. 
 
In particular, the embedding function in Eq.\eqref{embedding deformed sphere} should be extended by terms which are quadratic in the deformation parameter $\epsilon$, as given in Ref.\cite{Allais:2014ata}. For simplicity, we drop the $\sin(l\phi)$ type deformations. Thus, up to order $\mathcal{O}\left(\epsilon^2\right)$ we obtain
 
\begin{align}
    \Sigma_{\text{RT}}: \:\:r&=\sqrt{1-z^2}\bigg[1+\epsilon \sum_{l}\left(\frac{1-z}{1+z}\right)^{l/2}\left(\frac{1+lz}{1-z^2}\right)\cos{(l\phi)}\nonumber\\
    &+\epsilon^2\sum_{l}\left(\frac{1-z}{1+z}\right)^{l}\left(\frac{1}{4\left(1-z^{2}\right)^{2}}\right)\bigg(\left(1+2lz+\left(3l^{2}-2\right)z^{2}+2l\left(l^{2}-1\right)z^{3}\right)& \nonumber\\
    &+\left(2l\left(l^{2}-4\right)z^{3}+\left(3l^{2}-5\right)z^{2}+8lz+4\right)\cos\left(2l\phi\right)\bigg)+\mathcal{O}(\epsilon^4)\bigg].
    \label{embedding deformed sphere eps2}
\end{align}
In this formula, the explicit form of the term $R_{22}$ of Ref.\cite{Allais:2014ata} is presented, which to the best of our knowledge is a novel result. As a consistency check, we compute the expansion of the universal term coming from the RT formula \eqref{RTformula}, which reads

 \begin{equation}
    S_{\text{EH}}^{\text{Univ}}=-\frac{\pi\ell_{0}^{2}}{2G}\left[1+\epsilon^{2}\sum_l\frac{1}{4} l\left(l^{2}-1\right) -\epsilon^{4}\sum_l\frac{ l\left(23 l^{6}-246l^{4}+63 l^{2}-2\right)}{4\left(64l^{2}-16\right)} +\mathcal{O}(\epsilon^6)\right]\,.\label{EE deformed sphere for EH epsilon4}
\end{equation}
This formula, can equivalently be expressed solely in terms of the $C_{\text{T}}^{\text{EH}}$, given in \eqref{CT_EH}, as

 \begin{align}
    S_{\text{EH}}^{\text{Univ}}=&-\frac{\pi \ell_{0}^{2}}{2G}-\epsilon^2\frac{\pi^{4}C_{\text{T}}^{\text{EH}}}{24}\sum_ll\left(l^{2}-1\right)\nonumber\\ &+\epsilon^4\frac{\pi^{4}C_{\text{T}}^{\text{EH}}}{24}\sum_l\frac{ l\left(23 l^{6}-246l^{4}+63 l^{2}-2\right)}{16\left(4l^{2}-1\right)}+ \mathcal{O}(\epsilon^6) .\label{EE deformed sphere for EH epsilon4 in terms of CT}
\end{align}
Notice that in the quartic order term, only the $C_{T}$ coefficient appears. Even though we had anticipated that the higher-order coefficents $t_{2}$ and $t_{4}$ could appear generically, their absence is expected as they vanish for a CFT$_3$ dual to Einstein-AdS gravity. Furthermore, the last expression allows us to identify the geometry-dependent polynomial on the multipole moments $l$ multiplying the $C^{\text{EH}}_{\text{T}}$ at order $\mathcal{O} \left(\epsilon^{4}\right)$ that is 
\begin{equation}
    P_1(l)=\sum_l \frac{\pi^4 l\left(23 l^{6}-246l^{4}+63 l^{2}-2\right)}{384\left(4l^{2}-1\right)} \,.
    \label{P1}
\end{equation}
The relevance of this object will become clear in the analysis below.
 
For the case of CCG, the use of the RT surface in order to evaluate the HEE functional for the deformed sphere, is justified to the leading order in the higher curvature couplings, as discussed in Ref.\cite{Bueno:2020uxs}. Therefore, using the embedding given in Eq.\eqref{embedding deformed sphere eps2}, the terms of the HEE functional of \eqref{SR2}-\eqref{SK4 non-min} are given by

\begin{align}
     S_{\text{R}^2}&=\frac{6 \left(2 \mu_\mathrm{1}+4 \mu_\mathrm{2}+6 \mu_\mathrm{3}+24 \mu_\mathrm{4}+9 \mu_\mathrm{5}+9 \mu_\mathrm{6}+36 \mu_\mathrm{7}+144 \mu_\mathrm{8}\right)}{\ell_{\text{eff}}^{4}}\,,\\
     S_{\text{K}^2\text{R}}&=-\epsilon^2\frac{12z^{4}\left(\mu_\mathrm{1}-4 \mu_\mathrm{2}-2 \mu_\mathrm{3}-8 \mu_\mathrm{4}\right)}{\ell_{\text{eff}}^4\left(1-z^{2}\right)}\sum_{l}\left(\frac{1-z}{1+z}\right)^{l} l^{2} \left(l^{2}-1\right)^{2} \nonumber\\
     &-\epsilon^3\frac{12z^{4}\left(\mu_\mathrm{1}-4 \mu_\mathrm{2}-2 \mu_\mathrm{3}-8 \mu_\mathrm{4}\right)}{\ell_{\text{eff}}^{4}\left(1-z^{2}\right)^{3}}\sum_{l}\left(\frac{1-z}{1+z}\right)^{\frac{3 l}{2}}l^{2}\left(l^{2}-1\right) \left(2 l^{4} z^{2}-2 l^{3} z-14 l^{2} z^{2}\right.\nonumber\\
     &\left.+8 l^{2}+2 l z+3 z^{2}+1\right) \cos\left(l \phi\right)\nonumber\\
     &+\epsilon^4\frac{12\mathit{z}^{3}\left(\mu_\mathrm{1}-4 \mu_\mathrm{2}-2 \mu_\mathrm{3}-8 \mu_\mathrm{4}\right)}{\ell_{\text{eff}}^{4}\left(1-z^{2}\right)^{4} }\sum_{l}l\left(\frac{1-z}{1+z}\right)^{2 l}\bigg(\left(l^{2}-1\right)\big(l \left(6 l^{4}-23 l^{2}-1\right) z^{5}\nonumber\\
     &+3 \left(11 l^{4}-34 l^{2}+5\right) z^{4}+2 l \left(15 l^{4}-29 l^{2}-4\right) z^{3}+2l^{2}\left(14 l^{2}+4\right) z^{2} \nonumber\\
     &+l\left(53l^{2}-3\right) z-l^{4}+22 l^{2}-3\big) \cos\left(l \phi\right)^{4}+\frac{1}{2}\left(l^{2}-1\right)\big(2 l \left(14 l^{6}-88 l^{4}+111 l^{2}-16\right) z^{5}\nonumber\\
     &-\left(8 l^{6}+56 l^{4}-139 l^{2}+9\right) z^{4}+2 l \left(19 l^{4}-34 l^{2}+18\right) z^{3}-2 \left(5 l^{6}-5 l^{4}+24 l^{2}-3\right) z^{2}\nonumber\\
     &-2 l \left(2 l^{4}+17 l^{2}+5\right) z+2 l^{4}-29 l^{2}+3\big)\cos\! \left(l \phi\right)^{2}+\frac{l\left(1-z^{2}\right)}{4}\big(\big(8 l^{8}-112 l^{6}\nonumber\\
     &+332 l^{4}-168 l^{2}+21\big) z^{2}+4 l \left(4 l^{6}-17 l^{4}+22 l^{2}-9\right) z\nonumber\\
     &+8 l^{6}-142 l^{4}+68 l^{2}-15\big)\bigg)+O(\epsilon^5)\,,\\
     S_{\text{K}^4}^{\text{min}}&=-\epsilon^4\frac{3 z^{8}\left(\mu_\mathrm{1}-4 \mu_\mathrm{2}-2 \mu_\mathrm{3}-8 \mu_\mathrm{4}\right)}{\ell_{\text{eff}}^{4}\left(1-z^{2}\right)^{4}} \sum_{l}\left(\frac{1-z}{1+z}\right)^{2 l} l^{4} \left(l^2-1\right)^{4} +O(\epsilon^5)\,,\\
      S_{\text{K}^4}^{\text{non-min}}&=-\epsilon^4\frac{6 z^{8}\left(\mu_\mathrm{1}+2 \mu_\mathrm{2}\right)}{\ell_{\text{eff}}^{4}\left(1-z^{2}\right)^{4}} \sum_{l}\left(\frac{1-z}{1+z}\right)^{2 l} l^{4} \left(l^2-1\right)^{4}+O(\epsilon^5).
\end{align}
The difference that can be noted between the last two equations, makes manifest the splitting problem between the minimal and non-minimal prescriptions of the HEE functionals \eqref{HEE in CCG minimal Appendix} and \eqref{HEE in CCG nonminimal Appendix}, respectively. Summing up all the above contributions along with the corresponding counterterms, one notes two distinct terms contributing at order $\mathcal{O} \left(\epsilon^4\right)$. Nevertheless, one recognizes the presence of the polynomial $P_1(l)$ of Eq.\eqref{P1} found in the Einstein case, and by analogy, we identify its coefficient as the $C_{T}^{\text{CCG}}$, leading to the following expressions

\begin{align}
   S_{\text{CCG}}^{\text{Univ,min}}&=-\frac{\pi \ell_{\text{eff}}^{2}}{2G}C(\ell_{\mathrm{eff}},\vec{\mu})-\epsilon^2C_{\text{T}}^{\text{CCG}}\sum_l\frac{\pi^{4}}{24}l\left(l^{2}-1\right) \nonumber\\
   &+\epsilon^4\bigg(C_{\text{T}}^{\text{CCG}}  P_{1}\left(l\right)-\sum_l\frac{135 \pi l^{3} \left(l^{2}-1\right)^{3} \left(\mu_\mathrm{1}-4 \mu_\mathrm{2}-2 \mu_\mathrm{3}-8 \mu_\mathrm{4}\right)}{256 G \ell_{\text{eff}}^{2} \left(4 l^{2}-1\right) \left(4 l^{2}-9\right)}\bigg) +O(\epsilon^6) \,, \label{SCCGmin}\\
    S_{\text{CCG}}^{\text{Univ,non-min}}&=-\frac{\pi \ell_{\text{eff}}^{2}}{2G}C(\ell_{\mathrm{eff}},\vec{\mu})-\epsilon^2C_{\text{T}}^{\text{CCG}}\sum_l\frac{\pi^{4}}{24}l\left(l^{2}-1\right)  \nonumber\\
   &+\epsilon^4\bigg(C_{\text{T}}^{\text{CCG}} P_{1}\left(l\right)-\sum_l\frac{135 \pi l^{3} \left(l^{2}-1\right)^{3} \left(\mu_\mathrm{1}+2 \mu_\mathrm{2}\right)}{128 G \ell_{\text{eff}}^{2} \left(4 l^{2}-1\right) \left(4 l^{2}-9\right)}\bigg) +O(\epsilon^6) \,.
   \label{SCCGnonmin}
\end{align}
The fact that $P_1(l)$ appears in both EH and CCG theories indicates its theory independence within the HCG theories under consideration. Seeking similar quantities is fundamental in order to extract useful information out of the rest of the quartic order contribution. N.B. that $P_1(l)$ is not expected to remain theory independent for generic CFTs,
as here we are relating the (deformed) RT surface with the EE, what only
applies for theories that are dual to EH gravity plus perturbative corrections.

In the previous expressions, it is the coupling-constant dependence of the second term at order $\mathcal{O} \left(\epsilon^4\right)$ what makes evident the splitting problem. Interestingly enough, different holographic techniques indicate that in the presence of cubic curvature couplings, out of the two expressions for the quartic order term obtained considering the different splittings, only the non-minimal one has the coupling combination $(\mu_\mathrm{1}+2 \mu_\mathrm{2})$, present in the definition of the $t_4$ charge \cite{Sen:2014nfa,Chu:2016tps}. Therefore, the form of Eq.\eqref{SCCGnonmin} indicates that the quartic order in $\epsilon$ is given by a linear combination of $C_{T}$ and $t_{4}$, for CFTs dual to CCG.\footnote{For generic CFTs, the quartic order in the deformation parameter $\epsilon$ is not fully determined by $C_{T}$ and $t_{4}$, as shown by taking into account corner contibutions in Ref.\cite{Bueno:2015ofa}.}

In order to fix the normalization and determine $t_4$, we take into account the massless limit of CCG, studied in Ref.\cite{Li:2019auk}. Requiring the decoupling of the massive modes out of the particle spectrum of the theory leads, in $d=3$, to the equations

\begin{align}
    &12\mu_{7}+9\mu_{6}+5\mu_{5}+48\mu_{4}+16\mu_{3}+24\mu_{2}-3\mu_{1}=0 \,,\label{1st constranint}\\
    &432\mu_{8}+120\mu_{7}+36\mu_6+32\mu_{5}+16\mu_{4}+28\mu_{3}+6\mu_{1}+24\mu_{2}=0 \,.\label{second constraint}
\end{align}
Solving this system of equations, allows us to express the coupling constants $\mu_1$ and $\mu_2$ in terms of the rest of the couplings as

\begin{align}
    \mu_1&=-48\mu_8-12\mu_7-3\mu_6-3\mu_5-\frac{16}{3}\mu_{4}-\frac{4}{3}\mu_3 \,,\\
    \mu_2&=-6\mu_8-2\mu_7-\frac{3}{4}\mu_6-\frac{7}{12}\mu_5-\frac{8}{3}\mu_4-\frac{5}{6}\mu_3 \,.
\end{align}
Inserting these expressions into Eq.\eqref{SCCGnonmin}, one recovers the coupling dependence of the $t_4$ charge computed in Ref.\cite{Li:2019auk} in a splitting-independent way, namely

\begin{equation}
    t_4^{\text{CCG}-massless}=
   - \frac{120(360\mu_8+96\mu_7+27\mu_6+25\mu_5+64\mu_4+18\mu_3)}{\ell_{\text{eff}}^4}+\mathcal{O}(\mu^2) \,.
    \label{masslesst4}
\end{equation}
Unlike its non-minimal counterpart, the  minimal prescription \eqref{SCCGmin} does not match Eq.\eqref{masslesst4}, 
reafirming the inconsistency of the latter.

Determining the $t_4$ charge for the most generic CCG requires the identification of the geometry-dependent polynomial of the multipole momenta $P_{2} (l)$ that multiplies $t_4$. For the massless case, this polynomial adopts the form

\begin{equation}
     P_{2}(l)=\sum_l \frac{\pi^4}{2048}\frac{l^{3} \left(l^{2}-1\right)^{3}}{ \left(4 l^{2}-1\right) \left(4 l^{2}-9\right)} \,.
\end{equation}
Replacing the previous expression into Eqs.\eqref{SCCGmin} and \eqref{SCCGnonmin}, we determine the corresponding HEE functionals as

\small
\begin{align}
   S_{\text{CCG}}^{\text{Univ, min}}=&-\frac{\pi \ell_{\text{eff}}^{2}}{2G}C(\ell_{\mathrm{eff}},\vec{\mu})-\epsilon^2C_{\text{T}}^{\text{CCG}}\sum_l\frac{\pi^{4}}{24}l\left(l^{2}-1\right)\nonumber \\
   &+\epsilon^{4}C_{\text{T}}^{\text{CCG}}\left[P_{1}\left(l\right)-t_4^{CCG-min}P_{2}\left(l\right)\right] +\mathcal{O} \left(\epsilon^6 \right)\,,\\
    S_{\text{CCG}}^{ \text{Univ, non-min}}=&-\frac{\pi \ell_{\text{eff}}^{2}}{2G}C(\ell_{\mathrm{eff}},\vec{\mu})-\epsilon^2C_T^{CCG}\sum_l\frac{\pi^{4}}{24}l\left(l^{2}-1\right) \nonumber
    \\&+\epsilon^{4}C_{\text{T}}^{\text{CCG}}\left[P_{1}\left(l\right)-t_4^{CCG-non-min}P_{2}\left(l\right)\right] +\mathcal{O}\left(\epsilon^6 \right)\,,
    \label{SCCGfinal}
\end{align}
\normalsize
where the $t_4$ charges for both splittings become explicit and can be identified as

\begin{align}
    t_4^{\text{CCG-min}}&=\frac{360\left(\mu_\mathrm{1}-4 \mu_\mathrm{2}-2 \mu_\mathrm{3}-8 \mu_\mathrm{4}\right)}{\ell_{\text{eff}}^{4}\left(1+\frac{3}{\ell_{\text{eff}}^4}(4\mu_1-4\mu_2+2\mu_3+8\mu_4+9\mu_5+9\mu_6+36\mu_7+144\mu_8)\right)}\,,\label{t4 min}\\
   t_4^{\text{CCG-non-min}}&=
    \frac{720(\mu_1+2\mu_2)}{\ell_{\text{eff}}^4\left(1+\frac{3}{\ell_{\text{eff}}^4}(4\mu_1-4\mu_2+2\mu_3+8\mu_4+9\mu_5+9\mu_6+36\mu_7+144\mu_8)\right)}\,.\label{t4 nonmin}
\end{align}
Hence, it is the non-minimal prescription that recovers the value for $t_4$ obtained from the energy flux for graviton perturbations on a shockwave background, given by Sen and Sinha in Ref.\cite{Sen:2014nfa}.

We can also compare the expression of $t_4$ in the non-minimal splitting of Eq.\eqref{t4 nonmin}, with the corresponding expression for the case of Einsteinian Cubic Gravity (ECG) given in Ref.\cite{Bueno:2018xqc}. The definition of the theory considers

\begin{equation}
  \mu_1=-\frac{3}{2}\mu  \:\: ; \:\: \mu_2=-\frac{\mu}{8} \:\:  ;\:\:\mu_3= 0 \:\: ; \:\: \mu_4= 0 \:\: ; \:\:\mu_5= \frac{3}{2}\mu  \:\: ;\:\:  \mu_6= -\mu \:\: ;\:\:\mu_7= 0  \:\:;\:\:\mu_8=0\,.
\label{ECG_couplings}
\end{equation}
Therefore, one has

\begin{equation}
     t_4^{\text{ECG-non min}}=-\frac{1260\mu}{\ell_{\text{eff}}^{4}\bigg(1-\frac{3}{\ell_{\text{eff}}^4}\mu\bigg)}\,,
\end{equation}
in full agreement with this reference.

\section{Discussion}

In this work, the shape dependence of entanglement entropy is used as a tool in order to resolve the splitting problem in higher curvature theories of gravity. We begin our analysis by proving that the Kounterterm renormalization method can be used to isolate the universal part of the HEE in CFTs dual to generic HCG theories in $d\leq 4$. Then, CCG is considered as a toy model since it is the lowest-order HCG theory with manifest ambiguity in the definition of the HEE functional. In our resolution of the splitting problem, we promote the use of shape deformations of spherical entangling surfaces in order to compute the coefficients of stress tensor correlation functions. We use the splitting dependence of said coefficients in order to test the consistency of the minimal and non-minimal splittings.

We determine the $C$-function candidates that correspond to  the HEE for a spherical entangling surface in arbitrary dimensions, and then for the case of CFT$_4$, we also compute the type-$B$ anomaly. Afterwards, inducing sinusoidal  deformations of the spherical entangling surface, we are able to extract the coefficients $C_{\text{T}}^{\text{CCG}}$ and $t_4^{CCG}$ of the correlation functions of the CFT stress-tensor. The value of the coefficient $C_{\text{T}}^{\text{CCG}}$ matches the one given in the literature \cite{Bueno:2020uxs}. Most importantly, we show explicitly that this quantity is free of any ambiguity sourced by the splitting problem. 

However, this is not the case for $t_4^{CCG}$, as the results obtained in subsection \ref{subsection 3.3} for generic CCGs, exhibit explicit differences between the minimal and non-minimal splittings. The key point in the derivation, is the existence of polynomials that encode the multipole momenta of the sphere deformations and are independent of the theory in question. In particular,  we consider the CCG that carries solely massless graviton degrees of freedom \cite{Li:2019auk} in order to determine the polynomial in the Fourier expansion. Matching the results of the previous reference, which were computed in a splitting-independent way, allowed us to i) conclude that only the non-minimal prescription is consistent and ii) determine the normalization for the $t_4$ charge in the most generic CCG.  Also, it is worth emphasizing that our results for the non-minimal splitting match those of Ref.\cite{Bueno:2018xqc} for the particular case of ECG.

From the previous discussion, it becomes clear that $S^{\left(4\right)}$ contains information, namely $C_T^{\text{CCG}}$ and $t_4^{\text{CCG}}$, that determine the three-point correlators of the stress-energy tensor in CFT$_3$. The combination of the aforementioned charges at the quartic order of Eq.\eqref{SCCGfinal} resembles the structure of the one-point function of the energy flux $\langle{\mathcal{E} \left(\vec{n}\right)}\rangle$ \cite{Sen:2014nfa}, that arises in collider physics of CFTs \cite{Hofman:2008ar}. The presence of the two charges indicate that $S^{(4)}$ is the integrated version of a function that involves $\langle{TTT}\rangle$. In complete analogy, the energy flux of a state is defined as the integral of said correlator \cite{Hofman:2008ar,Buchel:2009sk}, making manifest the relation between $S^{(4)}$ and $\langle{\mathcal{E} \left(\vec{n}\right)}\rangle$. We conjecture, that the positivity of the latter can be understood as a consequence of the strong subadditivity of EE at $\mathcal{O}\left(\epsilon^4\right)$. The proof of this conjecture is out of the scope of the paper.

This work opens different avenues for future research. For example, the fact that the CFT stress-tensor correlator coefficients appear in the expansion of the renormalized HEE in terms of the deformation parameter, signals to a possible extension of the CHM map. In this extension, the HEE on the deformed spherical entangling surface could in turn be mapped to the partition function on a correspondingly deformed thermal sphere. It would be interesting to obtain the extended map between the distorted Euclidean-sphere manifold and the perturbed spherical entangling surface, in terms of their corresponding deformation parameters. Also, the fact that the renormalization scheme based on the extrinsic counterterms \eqref{HEE in CCG Kounterterm} was proven to work for isolating the universal part of the HEE in CFTs dual to generic HCGs, implies that the holographic study of CFT properties could be extended to the entire $\mathcal{L}(Riemann)$ class of gravity theories.

\acknowledgments

The work of GA is funded by ANID, Convocatoria Nacional Subvenci\'on a Instalaci\'on en la Academia Convocatoria A\~no 2021, Folio SA77210007. IJA and AA acknowledge funds from ANID, REC Convocatoria Nacional Subvenci\'on a Instalaci\'on en la Academia Convocatoria A\~no 2020, Folio PAI77200097. The work of RO is funded by FONDECYT Grant N$^{\circ }$1190533 {\it Black holes and asymptotic symmetries}, and ANID-SCIA-ANILLO Grant ACT210100 {\it Holography and its applications to High Energy Physics, Quantum Gravity and Condensed Matter Systems}.

\appendix

\section{Holographic Entanglement Entropy in Cubic Curvature Gravity} \label{Appendix A}

The HEE for the most general higher curvature Lagrangian  $\mathcal{L}_{\text{HCG}}(R_{\mu\nu\rho\sigma})$ is given by Dong formula in Eq.\eqref{Dong}. This expression, consists of two parts. The first one is nothing more than the usual Wald entropy formula which, in the context of black holes, generalizes the Bekenstein-Hawking entropy for a general theory of gravity. The second part involves corrections depending on the extrinsic curvature. The latter contribution has a more subtle origin and, as expected, it is not present in Einstein gravity.

When we regularize the cone singularity, we choose adapted coordinates around it. A parameter $\vartheta=2\pi\left(1-\frac{1}{n}\right)$ is introduced and an expansion of the curvature invariants is made with respect to said parameter. In principle, we should consider only terms up to linear order in $\vartheta$ due to the fact that higher orders would not contribute to the EE. In practice, what happens for higher curvature theories is that terms that --at first-- seem to be of second order $\mathcal{O}(\vartheta^2)$ can be enhanced to $\mathcal{O}(\vartheta)$ after the integration because of some would-be logarithmic divergence  when $\vartheta\rightarrow 0$. It is from this consideration that the anomalous term appears, correcting the Wald term.


The introduction of the anomalous term inherits an ambiguity in the determination of the corresponding HEE functional that is reflected in the summation over $\alpha$ in Eq.\eqref{Dong}. The presence of the second derivative of the Lagrangian with respect to the Riemann, indicates that this term must be a polynomial in the curvature tensors and it can be decomposed as $R_{A B i j}, R_{A i B j}$ and $R_{i j k l}$. These, in turn, are expanded in terms of the extrinsic curvature  $K^{A}_{ij}$, the codimension-two Riemann tensor $\mathcal{R}_{ijkl}$ and  the auxiliary tensors $\Tilde{R}_{A B i j}, \Tilde{R}_{A i B j}$  and $Q_{A B i j}$, whose definitions can be found in Ref.\cite{Dong:2013qoa}, as

\begin{align}
      R_{A B i j}&= \Tilde{R}_{A B i j}+g^{kl}[K_{A j k}K_{B i l}-K_{A i k}K_{B j l}],\label{minexpan R}\\
      R_{A i B j}&=\Tilde{R}_{A i B j}+g^{kl}K_{A j k }K_{A i l } -Q_{A B i j},\\
      R_{ijkl}&=\mathcal{R}_{ijkl}+G^{A B}[K_{A il}K_{B jk}-K_{A ij}K_{B kl}].\label{Minimalexpan}
\end{align}
The label $\alpha$ denotes each term in this expansion and $q_{\alpha}$ is a weight factor whose values are assigned accordingly as follows: we add one for each $Q_{zz ij}$ and $Q_{\bar{z} \bar{z} ij}$ and one half for each $K_{Aij}$,$R_{ABCi}$, $R_{Aijk}$; the number that we obtain for each term is $q_{\alpha}$. Then we divide each term by $\left(q_{\alpha}+1\right)$. After finishing the expansion and setting the weights for each term we can then use \eqref{minexpan R}-\eqref{Minimalexpan} to rewrite everything in terms of the original curvature tensor.

This procedure is dubbed as the minimal prescription and, as mentioned in section \ref{Section 2}, it is obtained without requiring the regularized metric at $\Sigma$ to satisfy the EOMs. In principle, it would be very difficult to impose this condition for an arbitrary HCG, however, it can be achieved for the case of the Einstein equations of motion \cite{Camps:2014voa,Miao:2015iba}. This has been shown to be consistent perturbatively in the higher curvature couplings and has been dubbed the non-minimal prescription.  Implementing the EOMs in the expansion of Eqs. \eqref{minexpan R}-\eqref{Minimalexpan} amounts to writing  the terms $R_{z\bar{z} z\bar{z}}$ and $Q_{z \bar{z} i j}$ as

\begin{align}
     R'_{z\bar{z}z\bar{z}}&=R_{z\bar{z}z\bar{z}}+\frac{1}{2}(K_{zij}K_{\bar{z}}^{ij}-K_z K_{\bar{z}}),\\
      Q'_{z\bar{z}ij}&=Q_{z\bar{z}ij}-K_{zi}^kK_{\bar{z}jk}-K_{zj}^kK_{\bar{z}ik}+\frac{1}{2}K_zK_{\bar{z}ij}+\frac{1}{2}K_{\bar{z}}K_{zij} \,.
\end{align}
In the non-minimal prescription one obtains the $q_{\alpha}$ by summing one half for each $K_{A ij}$, $R_{A B \gamma i}$ and  one for  each $Q_{zz ij}$ and $Q_{\bar{z} \bar{z} ij}$. As before, after expanding and assigning a weight $(q_{\alpha}+1)$ to each term one revert the expansion and express everything in terms of the curvature tensors. The fact that the regularization can be implemented in two different ways, leading to distinct results for the EE functional, is called the splitting problem. 

In Ref.\cite{Caceres:2020jrf} the explicit expressions for the two splittings were computed for the EE of  a holographic CFT dual to CCG \eqref{CubicLagrangian}. Using Eq.\eqref{Dong} and following the procedure for the minimal and non-minimal case, the authors obtained that

\begin{align}
  S_{\text{CCG}}^{\text{min}}&=\frac{\mathcal{A}[\Sigma]}{4 G}+\frac{1}{8 G}\int\limits_{\Sigma}d^{d-1}y\sqrt{|\sigma|}\left(S_{\text{R}^2}+S_{\text{K}^2\text{R}}+S_{\text{K}^4}^{\text{min}}\right), \label{HEE in CCG minimal Appendix}\\
  S_{\text{CCG}}^{\text{non-min}}&=\frac{\mathcal{A}[\Sigma]}{4 G}+\frac{1}{8 G}\int\limits_{\Sigma}d^{d-1}y\sqrt{|\sigma|}\left(S_{\text{R}^2}+S_{\text{K}^2\text{R}}+S_{\text{K}^4}^{\text{non-min}}\right),\label{HEE in CCG nonminimal Appendix}
\end{align}
where

\begin{align}
        S_{\text{R}^2}&=6\mu_8R^2+2\mu_7\left(R_{\mu\nu}R^{\mu\nu}+R_{A}^{\:\:\:A}R\right)+3\mu_6R_{A\mu}R^{A\mu}+\mu_5\left(2R_{\mu\nu}R_{A}^{\:\;\mu A\nu}-R_{AB}R^{AB}\right. \nonumber\\
        &\left.+R_{A}^{\:\:A}R_{B}^{\:\:B}\right)+2\mu_4\left(R_{\mu\nu\rho\sigma}R^{\mu\nu\rho\sigma}+2RR_{AB}^{\:\:\:\:AB}\right)+\mu_3\left(R_{A\mu\nu\rho}R^{A\mu\nu\rho}-4R_{A\mu}R_{B}^{\:\:A B\mu}\right) \nonumber\\
        &+6\mu_2R_{A B \mu \nu}R^{A B \mu \nu}-3\mu_1\left(R_{A\mu B\nu}R^{A \nu B\mu}-R_{A\mu\:\:\nu}^{\:\:\:\:\:A}R_{B}^{\:\:\mu B\nu}\right), \label{SR2} \\
        S_{\text{K}^2\text{R}}&=-\mu_7K_AK^AR-\frac{3}{2}\mu_6K_AK^AR_B^{\:\:B}-2\mu_5K_AK^A_{ij}R^{ij} 
        +\frac{1}{2}\mu_5K_AK^AR_{BC}^{\:\:\:\:BC} \nonumber\\
        &-4\mu_4K_{Aij}K^{Aij}R-2\mu_3K_{Aik}K^{A\:\:k}_{\:\:j}R^{ij}-\mu_3K_{Aij}K^{Aij}R_{B}^{\:B}-2\mu_3K_AK_{ij}^AR_B^{\:\:i Bj}\nonumber\\
        &-12\mu_2K_{Aik}K^{A\:\:k}_{\:\:j}R_{B}^{\:\:iBj}-3\mu_1K_{Aij}K^A_{\:\:kl}R^{ikjl}+\frac{3}{2}\mu_1K_{Aij}K^{Aij}R_{BC}^{\:\:\:\:BC}\nonumber\\
        &-6(2\mu_2+\mu_1)K_{Aik}K_{Bj}^{\:\:\:\:k}R^{ABij}, \label{SK2R} \\
      S_{\text{K}^4}^{\text{min}}&=-\frac{1}{2}\mu_7K_AK^AK_BK^B-\mu_5K_AK^A_{\:\:ij}\left(K_BK^{Bij}-2K_{B\:\:k}^{\:\:i}K^{Bjk}\right)
        +\frac{1}{4} \left(6\mu_7+3\mu_6 \right.\nonumber\\
        &\left.-8\mu_4\right)K_AK^AK_{Bij}K^{Bij}+\frac{1}{2}(12\mu_4+\mu_3-3\mu_1)K_{Aij}K^{Aij}K_{Bkl}K^{Bkl} \nonumber\\
        &+\frac{3}{2}(4\mu_2+3\mu_1)K_{Ai}^{\:\:\:\:j}K_{Bj}^{\:\:\:k}K^{A\:\:l}_{\:\:k}K^{B\:\:i}_{\:\:l}-(-2\mu_3+3\mu_1)K_{Ai}^{\:\:\:j}K^{A\:\:k}_{\:\:j}K_{Bk}^{\:\:\:l}K^{B\:\:i}_{\:\:l},\label{SK4 min}  \\
     S_{\text{K}^4}^{\text{non-min}}&=-\frac{1}{4}(\mu_5-3\mu_1)K_AK^AK_{Bij}K^{Bij}+\frac{1}{4}\mu_5K_AK^AK_BK^B \nonumber\\
        &-(\mu_3-6\mu_2)K_AK^A_{\:\:ij}K_{B\:\:k}^{\:\:i}K^{Bjk}+\mu_3K_AK^A_{\:\:ij}K_BK^{Bij} \nonumber\\
        &-\frac{3}{4}\mu_1K_{Aij}K^{Aij}K_{Bkl}K^{Bkl}-\frac{3}{2}\mu_1K_{Aij}K_{Bkl}K^{Bij}K^{Akl} \nonumber\\
        &+\frac{3}{2}(4\mu_2+3\mu_1)K_{Ai}^{\:\:\:\:j}K_{Bj}^{\:\:\:k}K^{A\:\:l}_{\:\:k}K^{B\:\:i}_{\:\:l}-3(4\mu_2+\mu_1)K_{Ai}^{\:\:\:j}K^{A\:\:k}_{\:\:j}K_{Bk}^{\:\:\:l}K^{B\:\:i}_{\:\:l}. \label{SK4 non-min}
\end{align}
Notice that the difference  between the two splittings is manifest in the term of quartic order in the extrinsic curvature, i.e., the anomalous term. On the other hand, the quadratic term in the curvature ($S_{\text{R}^2}$), is the same in both splittings, since its origin can be found in the Wald part, that is common for the two prescriptions. As for the last term ($S_{ \text{K}^2 \text{R}}$), it comes from the anomalous part but the counting procedure makes it splitting-independent.

\section{Renormalized  Cubic Curvature Gravity action}\label{Appendix B}

For an arbitrary HCG theory, the Kounterterms prescription requires the addition of the same boundary term as in the Einstein case, but with a different  coefficient $c^{\text{HCG}}_d$  which is fixed by requiring  a finite action for the pure AdS  vacuum. In this way, the method incorporates the information of the corresponding higher-curvature couplings, leading to the renormalized action given by

\begin{equation}
    I_{\mathrm{HCG}}^{ren}=\frac{1}{16\pi G}\int\limits_{\mathcal{B}}d^{d+1}x\sqrt{|G|}\mathcal{L}_{\mathrm{HCG}}+\frac{c^{\mathrm{HCG}}_d}{16\pi G}\int\limits_{\partial \mathcal{B}}d^{d}x\sqrt{|h|} B_d \,,\label{RenormalizedCubicCurvatureAction}
\end{equation}
where

\begin{equation}
c_d^{\text{HCG}}=c^{EH}_d C(\ell_{\text{eff}},\vec{\mu})
\label{cdHCG}
\end{equation}
with $c_d^{EH}$ given in Eq.\eqref{coupling Einstein Hilbert} and

\begin{align}
      C(\ell_{\mathrm{eff}},\vec{\mu})&=-\frac{\ell^2_{\text{eff}}}{2d}\mathcal{L}_{HCG}(\vec{\mu},\ell_{\mathrm{eff}})|_{AdS} \\
      &=\frac{\ell^3_{\mathrm{eff}}}{2d(d+1)}\frac{d\mathcal{L}_{\mathrm{HCG}(\vec{\mu},\ell_{\mathrm{eff}})}}{d\ell_{\mathrm{eff}}} \bigg|_{AdS} \,.\label{HCGCoupling}
\end{align}
Here,  $\mathcal{L}_{\text{HCG}}(\vec{\mu})|_{AdS}$ is the HCG Lagrangian evaluated on pure AdS, which depends on the vector of the higher-curvature couplings $\vec{\mu}$. The relation between the evaluated and differential forms of the above expression is given by Ref.\cite{Bueno:2016ypa}, and it allows to fix the coupling $C(\ell_{\mathrm{eff}},\vec{\mu}) $ without knowledge of the EOM of the theory.

The extrinsic counterterms are given in Refs.\cite{Olea:2005gb,Olea:2006vd} and adopt the form

\begin{equation}
   B_d =-(d+1)\int\limits_{0}^{1}\,dt\delta^{a_1...a_{d}}_{b_1...b_{d}}K^{b_1}_{a_1}\left(\frac{1}{2}\mathcal{R}_{a_2 a_3}^{b_2 b_3}-t^2K^{b_2}_{a_2}K^{b_3}_{a_3}\right) ...\left(\frac{1}{2}\mathcal{R}_{a_{d-1} a_{d}}^{b_{d-1} b_{d}}-t^2K^{b_{d-1}}_{a_{d-1}}K^{b_{d}}_{a_{d}}\right)  \,, \label{KTsodd}
   \end{equation}
for odd boundary dimension $d$  and

\begin{eqnarray}
   B_d &=& -d\int\limits_{0}^{1}dt\int\limits_{0}^{t}ds\,\delta^{a_1...a_{d-1}}_{b_1...b_{d-1}}K^{b_1}_{a_1}\left(\frac{1}{2}\mathcal{R}_{a_2 a_3}^{b_2 b_3}-t^2K^{b_2}_{a_2}K^{b_3}_{a_3}+\frac{s^2}{\ell_{\mathrm{eff}}^2}\delta^{b_2}_{a_2}\delta^{b_3}_{a_3}\right)\times ...\nonumber\\ &&...\times\left(\frac{1}{2}\mathcal{R}_{a_{d-2} a_{d-1}}^{b_{d-2} b_{d-1}}-t^2K^{b_{d-2}}_{a_{d-2}}K^{b_{d-1}}_{a_{d-1}}+\frac{s^2}{\ell_{\mathrm{eff}}^2}\delta^{b_{d-2}}_{a_{d-2}}\delta^{b_{d-1}}_{a_{d-1}}\right) \,,   \label{KTseven}  
\end{eqnarray}
for even $d$.

Here, $K_{a b}$  and $\mathcal{R}_{a b c f}$ are the extrinsic  and intrinsic curvatures of the AdS boundary $\partial  \mathcal{B}$, respectively. $\partial  \mathcal{B}$ belongs to the same conformal equivalence class as the CFT manifold $\mathcal{M}$.
Considering the coupling for the Kounterterm defined in Eq.\eqref{cdHCG}, the renormalized action for the HCG of Eq.\eqref{RenormalizedCubicCurvatureAction} becomes

\begin{equation}
    I_{\text{HCG}}^{\text{ren}}=\frac{1}{16\pi G} \mathcal{L}_{\text{HCG}}(\vec{\mu},\ell_{\mathrm{eff}})|_{AdS}\left(\mathrm{Vol}(\mathcal{B})-\frac{\ell_{\text{eff}}^2}{2d} c_{d}^{\text{EH}}\int\limits_{\partial \mathcal{B}}d^dx\sqrt{|h|} B_d\right) \,.
    \label{HCGren}
\end{equation}
The expression in parenthesis corresponds to the universal part of the AdS volume for asymptotically AdS manifolds whose dimension is lower or equal to five. For manifolds which are also asymptotically conformally flat, the expression is valid in any dimension. The universal part of the AdS volume is finite for even-dimensional manifolds or logarithmically-divergent for odd-dimensional ones \cite{Graham:1999jg,AST_1985__S131__95_0}.

\bibliographystyle{JHEP}
\bibliography{EELove}
\end{document}